\documentclass{aa}
\usepackage{psfig}
\usepackage{times}
\def\mag{\hbox{$^{\rm m}$}}
\begin{document}
\thesaurus{06                  
           (08.03.4;           
            08.16.5;           
            09.04.1;           
            09.08.1;           
            10.15.2 NGC 6383)} 
\headnote{Research Note}
\title{The central part of the young open cluster NGC 6383\thanks{
Based on observations collected at the European Southern Observatory, 
La Silla, Chile.}}

\author{M.E. van den Ancker\inst{1,2} \and P.S. Th\'e\inst{1} \and 
        D. de Winter\inst{1,3}}
\institute{
Astronomical Institute ``Anton Pannekoek'', University of Amsterdam, 
 Kruislaan 403, 1098 SJ  Amsterdam, The Netherlands \and
Harvard-Smithsonian Center for Astrophysics, 60 Garden Street, MS 42, 
 Cambridge  MA 02138, USA \and
TNO-TPD, P.O. Box 155, 2600 AD  Delft, The Netherlands}

\offprints{M.E. van den Ancker (mario@astro.uva.nl)}
\date{Received <date>; accepted <date>} 
 
\maketitle

\begin{abstract}
The spectral and extinction properties of 14 pre-main sequence candidates
in the central part of the very young open cluster NGC 6383 were investigated.
None of these stars shows evidence for anomalous
circumstellar extinction. However, six out of 14 programme stars do show
an infrared excess, indicative of the presence of circumstellar dust, heated
up by the central star. One of these stars (number 4), also shows H$\alpha$
in emission and shows some indications for the presence of circumstellar
gas in its spectrum, and might therefore be a newly identified Herbig Ae
star.
\end{abstract}

\section{Introduction} 
NGC 6383 ($\alpha_{1950}$ = 17$^{\rm h}$31\fm4, $\delta_{1950}$ = $-$32\degr32\arcmin) 
is a relatively small (10 arcmin radius, Hagen 1970) very young open cluster,
centered around the bright spectroscopic binary HD 159176 (O7 V + O7 V, 
$P$ = 3.366767 days; Stickland et al. 1993). The weak
H$\alpha$ emission nebula Stromlo 67 (Gum 1955), also known as S11 
(Sharpless 1953) or RCW 132 (Rodgers et al. 1960), is centered approximately
on HD 159176, by which it is excited. Together with NGC 6530 and NGC 6531,
NGC 6383 belongs to the Sgr OB1 association.

A photoelectric study by Eggen (1961) showed that the stars in NGC 6383 later
than A0 lie above the zero-age main sequence (ZAMS) and might therefore 
still be contracting. A photographic study of stars in a larger area 
around the cluster by Th\'e (1965) confirmed Eggen's results, but failed to 
find weak H$\alpha$ emission line stars of the type found in other very young 
open clusters, such as NGC 2264, NGC 6530 and NGC 6611.

Fitzgerald et al.
(1978) studied stars in the central core (1.25 arcmin radius) of NGC 6383
using photoelectric photometry and derived an average colour excess 
$E(B-V)$ = $0\fm33 \pm 0\fm02$, a cluster distance $d$ of $1.5 \pm 0.2$ kpc 
and a cluster age of $1.7 \pm 0.4$ million years. Luiken-Miller (1982) derived
a similar numbers for the $E(B-V)$: $0\fm31 \pm 0\fm02$. However, her cluster 
distance of $2.4 \pm 0.3$ kpc deviates considerably from the value by
Fitzgerald et al. (1978). Furthermore, Luiken-Miller found a considerable
spread in the ages of the stars in NGC 6383, with an average age of 2.2 
million years.

The spectral energy distributions of stars in the central core of NGC 6383 were
studied by Th\'e et al. (1985) using photoelectric photometry in the Walraven
$WULBV$, Cousins $VRI$ and Near-IR $JHKLM$ photometric systems. They found a
considerable infrared excess around three stars located above the ZAMS,
probably due to thermal emission of circumstellar dust grains, confirming the
pre-main sequence nature of these objects. Th\'e et al. also derived a colour
excess $E(B-V)$ = $0\fm30 \pm 0\fm01$ and a distance of $1.4 \pm 0.15$ kpc for
NGC 6383, in excellent agreement with the values found by Fitzgerald et al.
(1978).
\begin{table*}
\caption{Programme stars in NGC 6383.}
\begin{flushleft}  
\begin{tabular}{ccclcccl}
\hline\noalign{\smallskip}
\multicolumn{3}{c}{Star number} & Other  & \multicolumn{1}{c}{$\alpha$ (1950)} & 
\multicolumn{1}{c}{$\delta$ (1950)} & $V$ & Spectral\\
\cline{1-3}
F78 & T65 & E61 & designation & &            & [\mag] & type\\
\noalign{\smallskip}
\hline\noalign{\smallskip}
 ~1  & ~1  & ~1~  & HD 159176 & 17 31 26  & $-$32 32 57 & ~5.64 & O7 V + O7 V\\
 ~2  & ~2  & ~2~  & HD 317847 & 17 31 19  & $-$32 31 50 & 10.33 & B2 V\\
 ~3  & ~3  & ~3~  & HD 317857 & 17 31 18  & $-$32 34 12 & 10.30 & A1:IV:p\\
 ~4  & 27  & --   &           & 17 31 21  & $-$32 34 20 & 12.61 & A5 IIIp\\
 ~5  & 26  & --   &           & 17 31 15  & $-$32 32 39 & 12.86 & A2 Vep\\
 ~6  & --  & --   &           & 17 31 23  & $-$32 32 01 & 13.77 & A\\
 18  & 18  & 18~  &           & 17 31 20  & $-$32 33 26 & 13.37 & F8\\
 20  & 20  & 20~  &           & 17 31 22  & $-$32 31 51 & 11.42 & B9 IV\\
 21  & 21  & 21~  &           & 17 31 27  & $-$32 31 39 & 11.97 & F2 IV:p\\
 22  & 22  & 22~  &           & 17 31 33  & $-$32 33 21 & 12.30 & A3 V:e\\
 23  & 23  & 23~  &           & 17 31 31  & $-$32 32 22 & 13.79 & G3\\
 24  & 24  & 24~  &           & 17 31 32  & $-$32 33 14 & 11.35 & B9 Ve\\
 --  & 17  & 13a  &           & 17 31 16  & $-$32 31 27 & 12.54 & A3 Vp\\
 --  & 54  & --   &           & 17 31 40  & $-$32 33 20 & 12.32 & F0 Ve\\
\noalign{\smallskip}
\hline
\end{tabular}
\end{flushleft}
\end{table*}

However, spectral types of some of the stars studied by Th\'e et al. (1985) 
are still uncertain, and previous authors did not take the possibility of
anomalous extinction into account while deriving properties of the stars 
in NGC 6383. Therefore, a new study of the pre-main sequence candidates in
this cluster is in place. For this, 14 stars in NGC 6383, mostly located in
the core, were selected from the paper by Th\'e et al. (1985). The stars 
along with the available data in the literature are listed in Table~1. We 
shall use the star's number as given in Fitzgerald et al. (1978) except 
for the last two stars in Table~1 which are assigned number T17 and T54 
following the numbering system by Th\'e (1965).

\section{Observations}
\begin{table}
\caption{New $JHKLM$ photometry of star No. 6.}
\begin{flushleft}
\begin{tabular}{lccccc}
\hline\noalign{\smallskip}
Date    & $J$   & $H$  & $K$  & $L$  & $M$\\
\noalign{\smallskip}
\hline\noalign{\smallskip}
23/7/86 & 11.87 & 9.43 & 7.97 & 6.87 & 6.09\\
25/7/86 & 11.98 & 9.46 & 7.97 & 6.85 & 6.53\\
\noalign{\smallskip}
\hline\noalign{\smallskip}
Error   & ~0.02 & 0.01 & 0.01 & 0.08 & 0.42\\
\noalign{\smallskip}
\hline
\end{tabular}
\end{flushleft}
\end{table}
\begin{figure}
\vspace{0.2cm}
\centerline{\psfig{figure=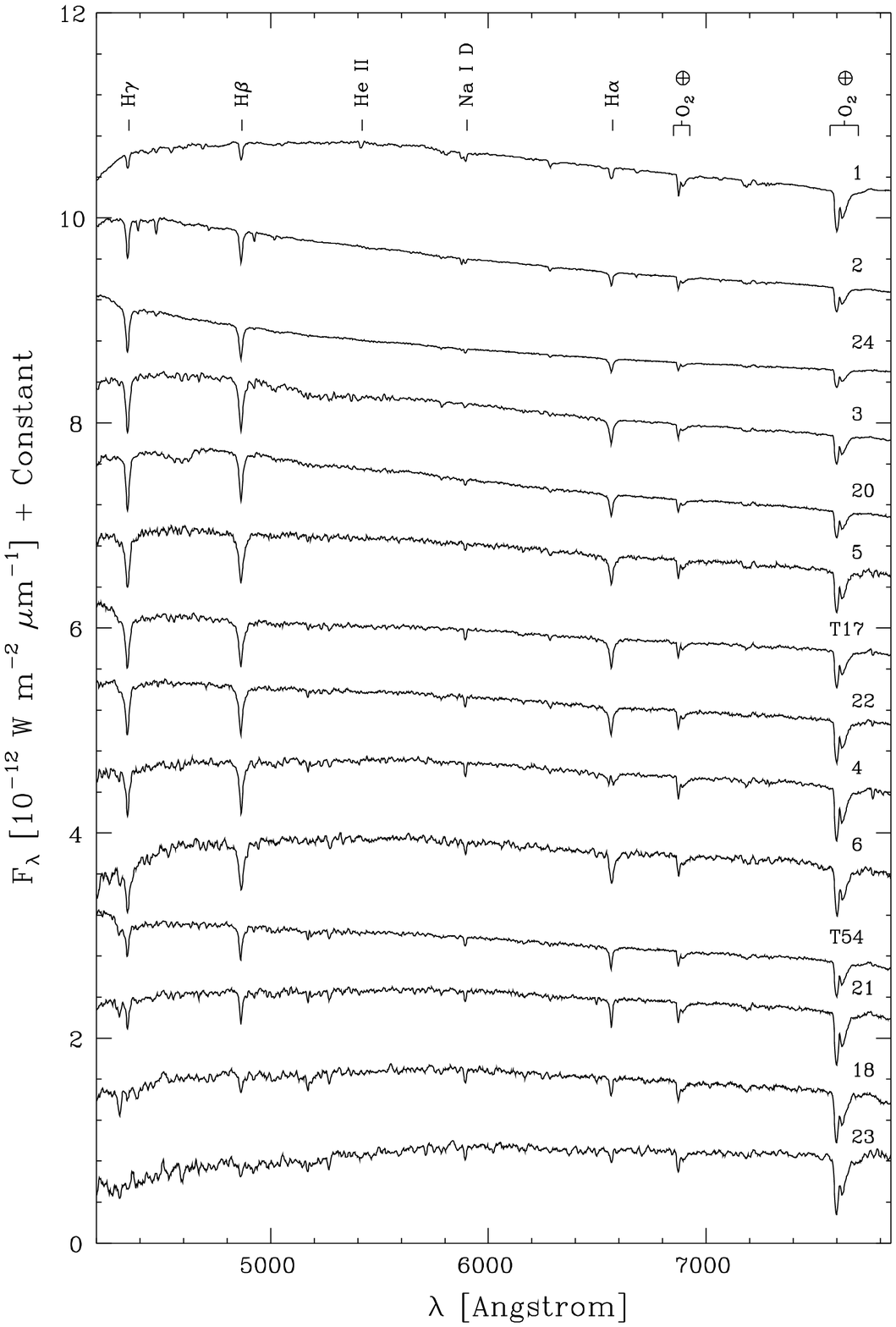,width=8.4cm}}
\caption[]{Low-resolution spectra of programme stars in NGC 6383 taken with
the ESO 1.5 m.}
\end{figure}
Low resolution (1.80~\AA~pix$^{-1}$) CCD spectra of our programme stars, in 
the wavelength range of
4200 to 7800~\AA, were obtained on the 16$^{\rm th}$ of June 1992 at the
European Southern Observatory, La Silla, Chile, using the Boller \& Chivens
spectrograph mounted on the ESO 1.52 m telescope. Using the MIDAS software 
package, the spectra were reduced with the usual steps of bias subtraction, 
flatfielding, background subtraction and spectral extraction, and wavelength 
and flux calibration. The resulting spectra are shown in Fig.~1.

New $JHKLM$ near-IR photometric data of star number 6 in the numbering
system by Fitzgerald et al. (1978) were obtained by P.S. Th\'e in July 1986
with the ESO 1 m telescope at La Silla with the InSb detector attached. These
observations were made through a 15\arcsec\ diaphragm. Sky subtraction was
achieved by chopping, with a frequency of 8~Hz, in the east-west direction
with a throw of 30\arcsec\ amplitude. The measurements and their accuracies
are listed in Table~2.
\begin{figure*}
\vspace{0.3cm}
\centerline{\psfig{figure=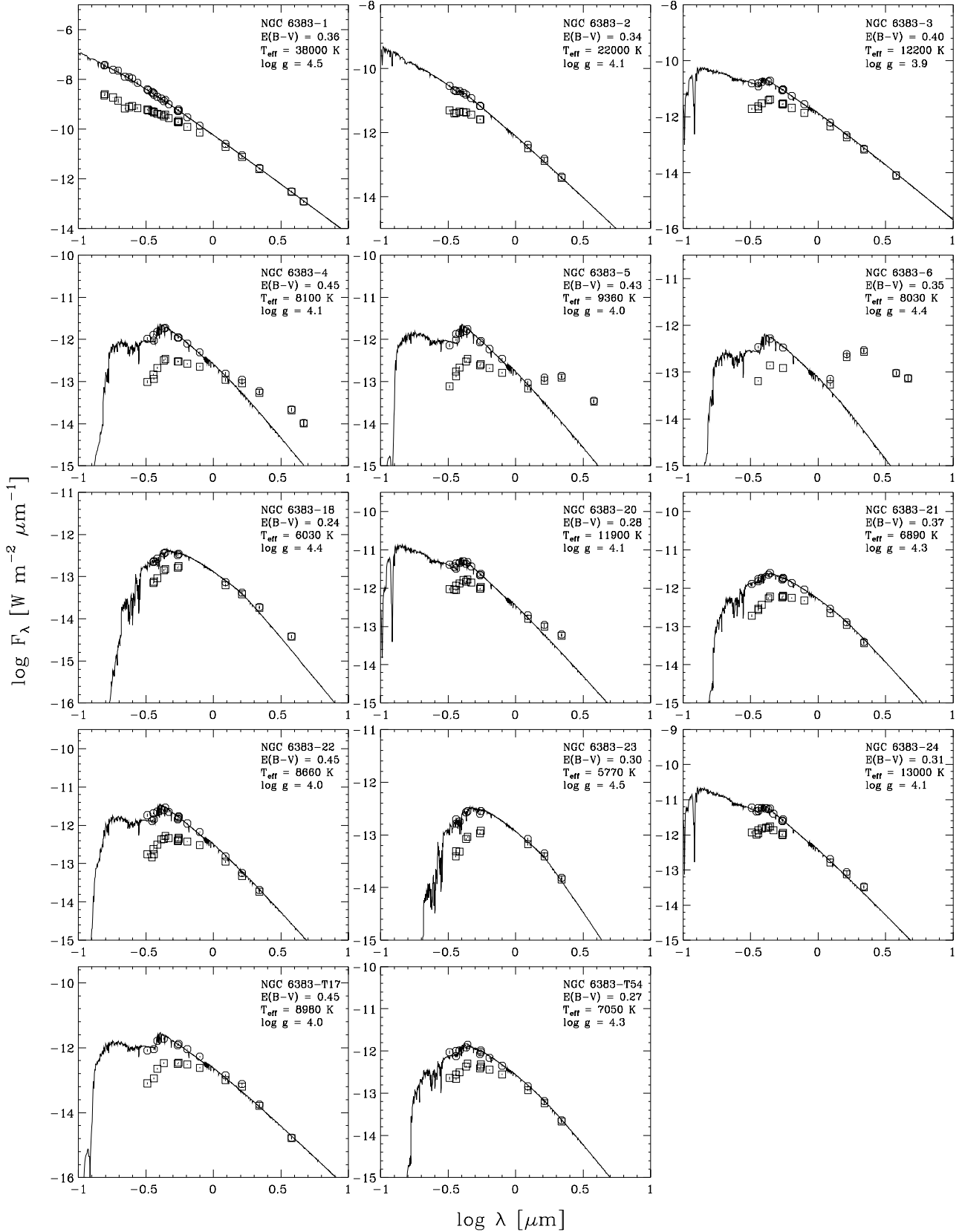,height=22cm}}
\caption[]{Observed (squares) and extinction-free (circles) SEDs for all
programme stars. The solid lines are the Kurucz (1991) models fitted through the
extinction-free SEDs.}
\end{figure*}

\section{Spectral classification}
New spectral classifications for our programme stars were
made by comparing our new spectra with the standard MK spectra by Jacoby et al.
(1984). The resulting spectral classifications are listed in Table~3.
Particularly interesting is the spectrum of star number 4, which shows
H$\alpha$ in emission and also has the O\,{\sc i} 7774 line strongly in
absorption. According to the relations obtained by Danks \& Dennefeld (1994),
the relative strength of this last line indicates that it must be
a giant. Furthermore, the Na\,{\sc i}\,$D$ lines in this star appear to be
somewhat stronger than those in the other spectra shown in Fig.~1, suggesting 
the presence of circumstellar gas in this object.

Additional spectral classifications were made from the photometric data by
Fitzgerald et al. (1978) and by Eggen (1961) by deredenning all programme
stars in the $(U-B)$ versus $(B-V)$ diagram, assuming a normal
extinction law, i.e. $E(U-B)/E(B-V)$ = 0.72. They are listed in Table~3. 
The adopted final classification, determined by comparing literature 
classifications with ours is also listed in Table~3. Note that with 
the exception of star number 4 we do not retrieve the emission-line 
classification obtained for several programme stars by previous 
authors. We believe this to be due to the superior background subtraction 
procedures allowed by our long-slit CCD observations.
\begin{table*}
\caption{Astrophysical parameters of programme stars.}
\begin{flushleft}
\tabcolsep0.12cm
\begin{tabular}{rlllccccc}
\hline\noalign{\smallskip}
Star & \multicolumn{3}{c}{Spectral type}     & $E(B-V)$ & $R_V$ & $\log T_{eff}$ & $\log L_\star/L_\odot$ & IR Excess\\
\cline{2-4}
     & Photometric & Spectroscopic & Adopted &          &       &                & & \\
\noalign{\smallskip}
\hline\noalign{\smallskip}
1    & O9        & O7       & O7 V    & 0.36   & 3.1  & 4.580 & 5.777 & N\\
2    & B2        & B2       & B2 V    & 0.34   & 3.1  & 4.342 & 3.324 & N\\
3    & B9/A8     & B8       & B8 IV   & 0.40   & 3.1  & 4.086 & 2.858 & N\\
4    & A0/A8     & A5 IIIe  & A5 IIIe & 0.45   & 3.1  & 3.908 & 1.695 & Y\\
5    & B9/A8     & A1 IV    & A1 IV   & 0.43   & 3.1  & 3.971 & 1.659 & Y\\
6    & B9/A8     & A6       & A6 V    & 0.35   & --   & 3.905 & 1.166 & Y\\
18   & B7/F5/G5  & G0       & G0 V    & 0.24   & 3.1  & 3.780 & 1.194 & Y\\
20   & B8        & B7       & B8 V    & 0.28   & 3.1  & 4.076 & 2.243 & Y\\
21   & B8/F2/G5  & F2       & F2 V    & 0.37   & 3.1  & 3.838 & 1.884 & N\\
22   & A0/A7     & A3 IV    & A3 IV   & 0.45   & 3.1  & 3.938 & 1.850 & N\\
23   & B7/F5/G5  & G5       & G5 V    & 0.30   & 3.1  & 3.761 & 1.121 & N\\
24   & B6        & B7       & B7 V    & 0.31   & 3.1  & 4.114 & 2.383 & Y:\\
T17  & A0        & A2 IV    & A2 IV   & 0.45   & 3.1  & 3.953 & 1.752 & N\\
T54  & B7/F5     & F1       & F1 V    & 0.27   & 3.1  & 3.848 & 1.614 & N\\
\noalign{\smallskip}
\hline
\end{tabular}
\end{flushleft}
\end{table*}

\section{Spectral Energy Distributions}
Spectral energy distributions (SEDs) of all programme stars were constructed
using photometric data from Th\'e et al. (1985), Fitzgerald et al. (1978),
Eggen (1978), the ANS catalogue (Wesselius et al. 1982), the TD1 catalogue
(Thompson et al. 1978) and our new photometry from Table~2. No reliable
IRAS data were available for any of our programme stars. The resulting SEDs,
shown in Fig.~2, were analyzed by comparing them to Kurucz (1991)
models corresponding to their spectral classification, using the method
described by van den Ancker et al. (1997). The values for the ratio of 
total to selective extinction $R_V$, which are the 
result of the application of this procedure are also given in Table~3. No 
deviation from a normal interstellar extinction law (i.e. $R_V$ = 3.1) 
could be found in any of these stars.

As can be seen from Fig.~2, three of our programme stars (numbers 4, 5 and 6)
show large amounts of excess radiation above photospheric levels, similar to
those seen in many Herbig Ae/Be stars, in the near-infrared. Two additional
stars (numbers 18, 20 and possibly 24) show a modest amount of infrared 
excess. This probably indicates the presence of circumstellar dust, heated up 
by the central star, in these systems. Since star number 4 is the only one of
our programme stars which also shows H$\alpha$ in emission (see Fig.~1),
this makes it a new candidate for membership of the Herbig Ae/Be
stellar group in our sample.

\section{Evolutionary status}
\begin{figure}
\vspace{0.3cm}
\centerline{\psfig{figure=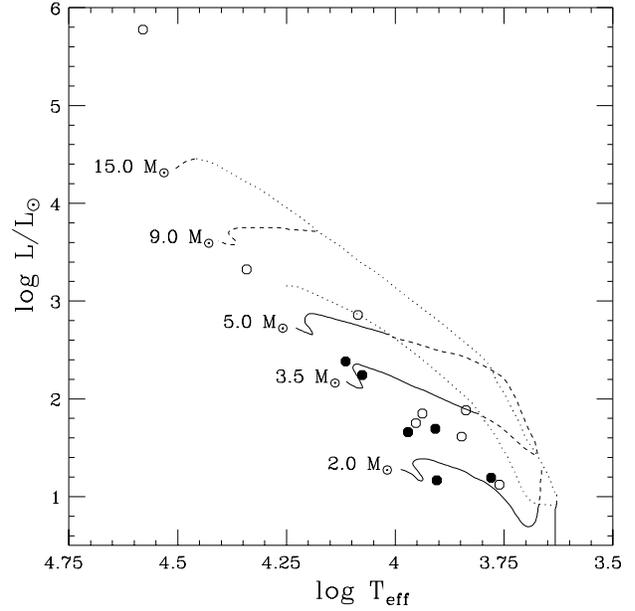,height=8cm,angle=90}}
\caption[]{Hertzsprung-Russell diagram of our programme stars in NGC 6383.
The filled circles indicate the positions of the stars with infrared excess. 
Also shown are the theoretical pre-main sequence evolutionary tracks (solid
lines and dashed lines) and the birthlines for 10$^{-4}$ (upper dotted line) 
and 10$^{-5}$ M$_\odot$~yr$^{-1}$ (lower dotted line) by Palla \& Stahler 
(1993).}
\end{figure}
Again following the method of van den Ancker et al. (1997), luminosities of 
our programme stars were computed from their SEDs, adopting the distance of 
$1.4 \pm 0.15$ kpc towards NGC 6383 obtained by Th\'e et al. (1985). The 
computed luminosities are also listed in Table~3.
Using these and the temperatures corresponding to the spectral
types, a Hertzsprung-Russell diagram was constructed (Fig.~3).
In this diagram we also plotted the pre-main sequence evolutionary tracks by
Palla \& Stahler (1993), as well as their so-called ``birthline'', where a
star becomes visible for the first time during its evolution towards the
main-sequence, for accretion rates of 10$^{-5}$ and 10$^{-4}$ 
M$_\odot$~yr$^{-1}$.

From Fig.~3 we notice that all stars in NGC 6383 fall below both stellar
birthlines, and therefore confirm the Palla \& Stahler models. Furthermore, 
the stars we classified spectroscopically with 
luminosity classes III or IV in Table~3 all seem to be located to the 
right of the main-sequence, and therefore are probably true pre-main sequence 
stars. However, no strong correlation between position in the HR diagram and 
the presence of a near-infrared excess seems to be present within our sample.

\section{Conclusions}
The spectral and extinction properties of 14 pre-main sequence star candidates
in the central part of the very young open cluster NGC 6383 were investigated.
From fitting of Kurucz (1991) models to their Spectral Energy Distributions,
it was concluded that none of these stars shows any evidence for anomalous
circumstellar extinction. However, six out of out 14 programme stars do show
an infrared excess, indicative of the presence of circumstellar dust, heated
up by the central star. No correlation between the infrared excess and the 
position in the cluster's HR diagram could be found.

One programme star (number 4), has a large infrared excess, reminiscent of 
the one displayed by many Herbig Ae/Be stars. Furthermore, it shows H$\alpha$ 
in emission and shows some indications for the presence of circumstellar
gas in its spectrum. Therefore it was concluded that this star can be 
considered as a new Herbig Ae star candidate. A literature search was done 
for photometric data for this star, but the available data do not show any 
indications for photometric variability, such as displayed by many Herbig Ae 
stars.

\acknowledgements{The authors would like to thank an anonymous referee for 
useful comments. This research has made use of the Simbad data base, operated 
at CDS, Strasbourg, France.}

\end{document}